\documentclass[11pt,a4paper,onecolumn]{article}
\usepackage{graphicx}
\usepackage{hyperref}
\usepackage{float}
\usepackage{caption}
\usepackage{lipsum}
\usepackage{amsmath}

\begin{document}

\title{NMSSM Inflation and Domain Walls from a Tri-critical Point of View}

\author{H. Gholian Aval \\
Physics Department, University of Tehran\\
\texttt{hadi.gholian@alumni.ut.ac.ir}}

\date{March, 2017}

\maketitle

\begin{abstract}
 We study the conditions in which we could bring a universe filled with different $Z_3$ vacua created during the Next to Minimal Supersymmetric Standard Model (NMSSM) electroweak symmetry breaking at $\textit{O} (10)^2$ GeV and a three dimensional three states diluted Potts model together in the same universality class. Then we use Cardy-Jacobsen conjecture to prove that there might be a tri-critical point in the NMSSM electroweak epoch of early universe. We prove that due to the existence of this point there would be no cosmological domain wall problem. Moreover, at this point the heat capacity and correlation length diverge which lead to a huge amount of energy release at constant temperature and a new mechanism for cosmological structure formation. Also, the entropy decrease after the tri-critical phase transition could explain the problem of low initial entropy in early universe. Finally, we combine Cardy-Jacobsen and Yaffe-Svetitsky conjectures to show that there might be a tri-critical point for the $Z_3$ center symmetry breaking in a pure SU(3) theory at $\textit{O} (10)^2$ MeV. Near the pure SU(3) tri-critical point, unlike the electroweak epoch tri-critical point, entropy increases rather than decrease.
\end{abstract}

\section{Introduction}
\label{sec.1}

Next to Minimal Supersymmetric Standard Model (NMSSM) is an extension of Minimal Supersymmetric Standard Model (MSSM) to solve the $\mu$-problem \cite{c.1,c.2}. According to Kibble-Zurek mechanism \cite{c.3,c.4,c.5}, the problematic cosmological domain walls which are topological defects would be created during the spontaneous discrete $Z_3$ electroweak symmetry breaking at $\textit{O} (10)^2$ GeV. It was first noted by Zel'dovic, Kobzarev and Okun in 1975 \cite{c.6} that if discrete symmetries of scalar field theory are spontaneously broken as the universe cools down, there would be further drastic problems in its evolution. If such domains remain they can over close the universe by dominating the energy density. It has been shown that these defects can cause huge amount of anisotropies in the CMB radiation, and to avoid this problem their energy scale must be less than a few MeV \cite{c.7}. Actually to be more concise it must be less than 0.93 MeV \cite{c.8}. As an indispensable consequence, these problematic walls must be lifted before the nucleosynthesis era.

It was first suggested by Zel'dovic \textit{et al.} in \cite{c.6} that breaking the degeneracy of the vacua, eventually leads to the dominance of true vaccum. In this case, the difference in energy density between the distinct vacua begins to exceed the tension which leads to unstable domain wall \cite{c.9}. In this paper it is shown that the vacua degeneracy is impossible due to the fact that the random percolation threshold in three dimensions is less than $\frac{1}{3}$ \cite{c.10,c.11} (for a review about percolation theory and its application see \cite{c.12}).

In this paper we want to look at the cosmological domain wall problem from a new perspective. In section (\ref{sec.2}), we study the conditions in which we could bring a universe filled with different $Z_3$ vacua and a three dimensional three states Potts model together in the same universality class. We use Cardy-Jacobsen (C-J) conjecture \cite{c.13} to prove that there might exists a tri-critical point in the NMSSM electroweak symmetry breaking in the early universe. Then in sections (\ref{sec.3}) and (\ref{sec.4}), we use the Landau theory to investigate the properties of a tri-critical phase transition in order to use them in early cosmology, i.e. an infinite energy release at constant temperature due to the divergence of heat capacity, or the divergence of correlation length which leads to a new mechanism for the cosmological structure formation and electroweak inhomogeneity problem. In this mechanism, cosmological domain walls cause no problem and vanish at the tri-critical point, also there might be a late-time inflation due to trapping in a metastable state, which had been suggested and studied in \cite{c.14,c.15,c.16,c.17,c.18} to solve the thermal and non-thermal overproduction in the early universe.

In section (\ref{sec.5}), we combine these results with Yaffe-Svetitsky (Y-S) conjecture \cite{c.19}, to prove that there might exists a tri-critical point during the $Z_3$ center symmetry breaking in a pure SU(3) theory too. We study the cosmological consequences of this point at $\textit{O} (10)^2$ MeV. At these energy scales we can neglect quarks as in quenched (static quarks) approximation of QCD, and there would be a pseudo-inflation mechanism in which without any exponential expansion of scale factor or further reheating, we have huge energy and entropy release.

\section{Cardy-Jacobsen Conjecture}
\label{sec.2}

C-J conjecture predicts the effects of quenched disorder on systems that undergo a $1st$ order phase transition in the ideal pure condition \cite{c.13}. According to this conjecture upon dilution, a $1st$ order phase transition in a three dimensional three states Potts model changes to a $2nd$ order at a tri-critical point \cite{c.13,c.20}.

We want to use the number of randomly diluted sites together with the total surface area of domain clusters in order to present a relation for the thickness of domain clusters walls in a Potts model with the nearest neighbors interaction. Hamiltonian of a non-diluted Potts model is \cite{c.21}:

\begin{equation}\label{eq.1}
\mathcal{H}=-\sum_{r,r^\prime}\,\, \delta_{\sigma_r , \,\sigma_{r^\prime}}
\end{equation}

\begin{figure}[b!]
  \centering
  \includegraphics[width=1\textwidth]{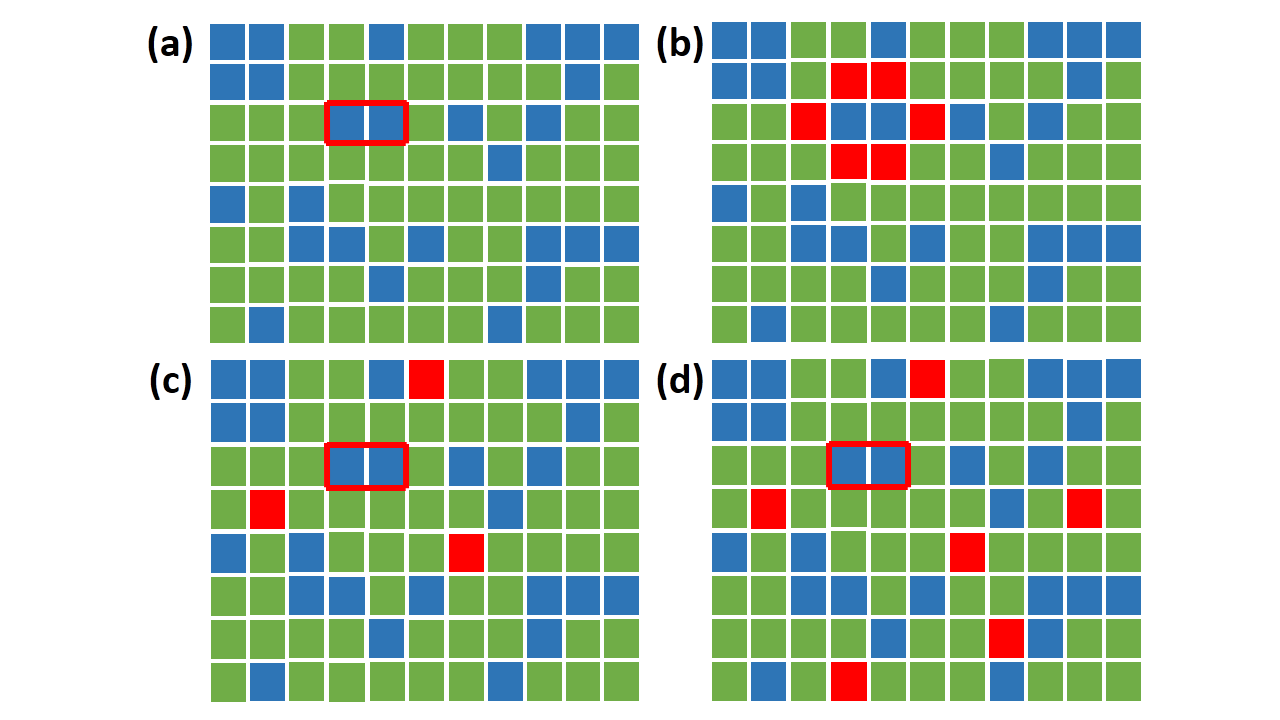}
  \captionsetup{width=1\textwidth}
  \caption{\small {(a) A typical two-site domain cluster with walls of zero thickness. (b) Diluted neighbors and non-zero thickness which leads to serious problems in the numerical simulations. (c) Dividing the number of randomly diluted sites by the surface area can give us a non-zero equivalent thickness $\Delta\ell$. (d) More diluted sites means thicker walls.}}\label{Fig.1}
\end{figure}

Here $\sigma_r$ is the spin state of site $r$ with $r^\prime$s as its nearest neighbors. For simplicity, we consider a typical two states Potts model in two dimensions Fig. (\ref{Fig.1}). In this figure green and blue sites represent up and down spin states. Thin surface of the two-site blue domain cluster is displayed by red lines in Fig. (\ref{Fig.1}.a). Similarly, for other domain clusters we have these walls too. The problem is that walls thickness are not taken into account in the numerical lattice simulations.

So the question is how can we take the walls thickness into our numerical simulations?  As we see in Fig. (\ref{Fig.1}.b), a probable way is to dilute the neighbor sites of the cluster and use diluted Potts Hamiltonian

\begin{equation}\label{eq.2}
\mathcal{H}=-\sum_{r,r^\prime}\varepsilon_r \varepsilon_{r^\prime} \,\, \delta_{\sigma_r , \,\sigma_{r^\prime}}
\end{equation}

Here each spin interacts with its nearest neighbors, and $\varepsilon_r$ is occupation number which is zero for each neighbor site, else it is equal to one. According to this Hamiltonian, if one of the neighbors of site $r$ is diluted then its contribution would be zero. As a consequence, this configuration in Fig. (\ref{Fig.1}.b) leads to further problems in the evolution of two-site blue cluster with nearest neighbors interaction. Due to the existence of diluted sites in the vicinity, this cluster is isolated from the rest of system. If we have these walls for other clusters, there won't be any evolution in the system by changing the temperature.

In order to avoid this problem, we have two solutions: first, use Potts model with a next to nearest neighbors Hamiltonian. This solution might be useful for the simulations in the condensed matter realm, but not for the purposes of this paper. Here we want to make a connection between a $3d$ three states diluted Potts model and an early universe filled with three different $Z_3$ vacua created during an NMSSM electroweak symmetry breaking. Actually our final goal is to bring them together in the same universality class.

\begin{figure}[t!]
  \centering
  \frame{\includegraphics[width=1\textwidth]{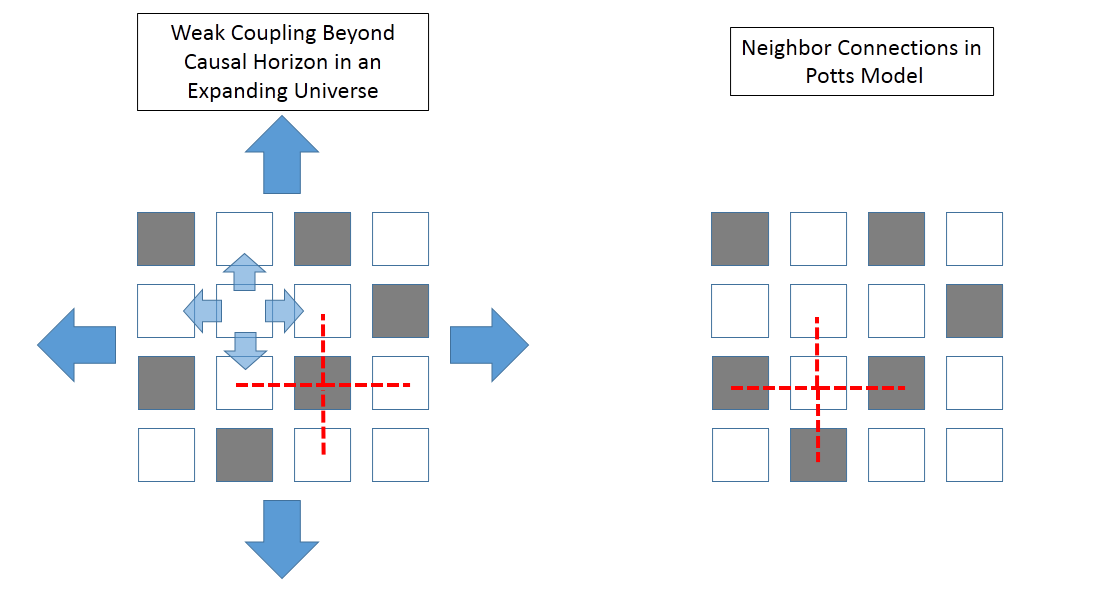}}
  \captionsetup{width=1\textwidth}
  \caption{\small {(Left) An expanding universe filled with regions occupied by different $Z_3$ vacua created during the electroweak symmetry breaking in NMSSM. Each site is a region restricted with its causal horizon. The red lines show a connection beyond the horizons. (Right) A Potts model with nearest neighbors interaction.}}\label{Fig.2}
\end{figure}

There exists a $Z_3$ symmetry associated to the Hamiltonian of these two systems in three spatial dimensions. For the NMSSM electroweak vacua, we will see in the next sections that there might be a possibility for the neighbors to be connected beyond their causal horizon (Fig. (\ref{Fig.2})), similar to the nearest neighbors interaction in the Potts model.

With these three conditions, i.e. same symmetry of Hamiltonian, number of spatial dimension, and range of interactions, no matter whether the universe expands or not, these two systems belong to the same universality class. As a result, near the (tri)critical point we can study the properties of a diluted Potts model and use them for the NMSSM electroweak symmetry breaking with domain walls in the early universe.

The next probable solution to take the walls thickness into the numerical simulations would be random dilution \cite{c.13,c.20,c.22}. For this case as we see in Fig. (\ref{Fig.1}.c), we randomly dilute some of the sites in our system. In order to make a relation between random dilution and domain walls structure, we define equivalent thickness $\Delta\ell$ as

\begin{equation}\label{eq.3}
\Delta\ell=\frac{N}{A}
\end{equation}

Here $\Delta\ell$ would be equivalent thickness of domain walls in the system if $N$ is total number of diluted sites, and $A$ is total surface area of domain clusters. We must pay attention that there is no thick walls in the random diluted Potts model. In other words, \emph{Potts model with $N$ random diluted sites and total domain clusters of surface area $A$ with zero thickness in numerical simulations, is equal to a Potts model with domain clusters walls of thickness $\Delta\ell$} . This simple statement would be an extremely helpful tool to make a connection between diluted sites in Potts model and domain walls in the NMSSM electroweak symmetry breaking to solve the cosmological domain walls problem.

\begin{figure}[t!]
  \centering
  \includegraphics[width=1\textwidth]{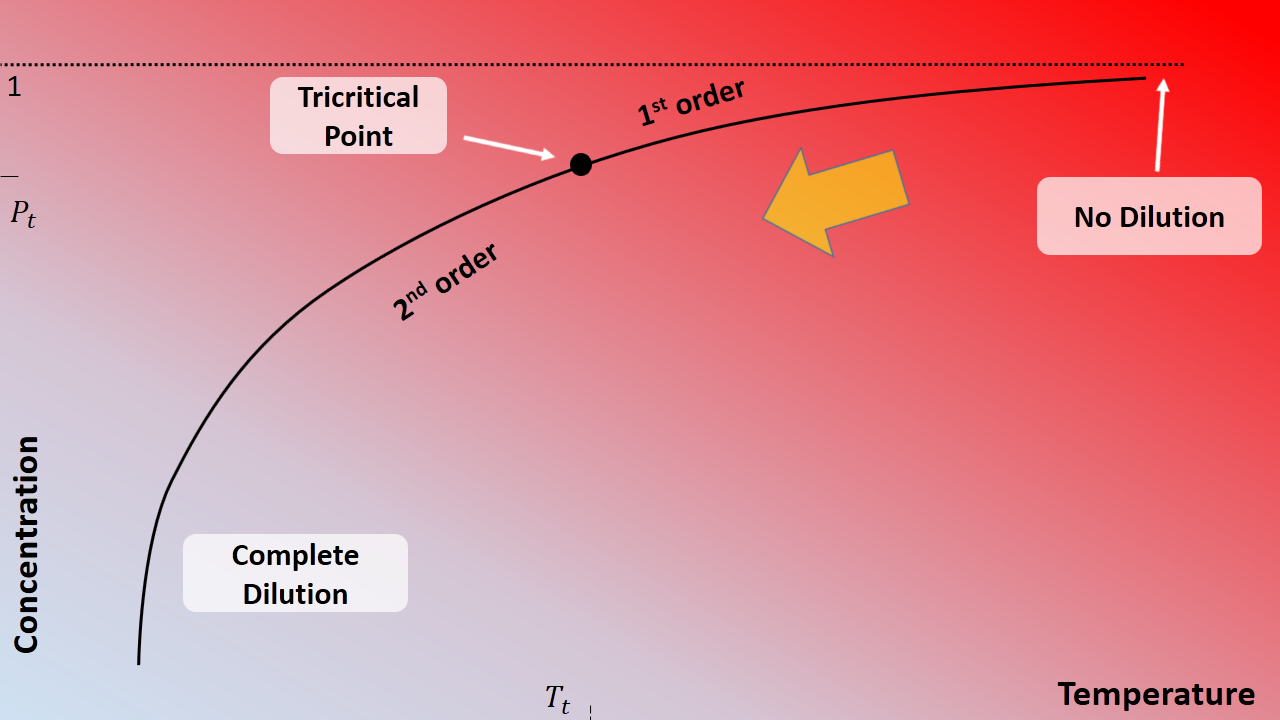}
  \captionsetup{width=1\textwidth}
  \caption{\small {Schematic description of the phase transition evolution. When temperature decreases, the order of phase transition changes at a tri-critical point; figure adopted from Ref. \cite{c.22}. }}\label{Fig.3}
\end{figure}

Again we consider the two-site blue cluster in Fig. (\ref{Fig.1}) (for a complete calculation we must take all the domain clusters surfaces into account). According to Eq.(\ref{eq.3}), the equivalent wall thickness for this domain cluster in Fig. (\ref{Fig.1}.c) and Fig. (\ref{Fig.1}.d) is

\begin{equation}\label{eq.3.1}
\Delta\ell_{(c)}=\frac{3}{6} \qquad \qquad \Delta\ell_{(d)}=\frac{6}{6}
\end{equation}

Eq. (\ref{eq.3}) suggests that an increase in the diluted sites is equal to an increase in the equivalent thickness of the walls. The subtle point is that near a tri-critical point in the evolution of a diluted Potts model, the total surface area of domain walls diverges which leads to a zero equivalent thickness $\Delta\ell$. Later we will study this in more details.

\begin{figure}[t!]
  \centering
  \frame{\includegraphics[width=1\textwidth]{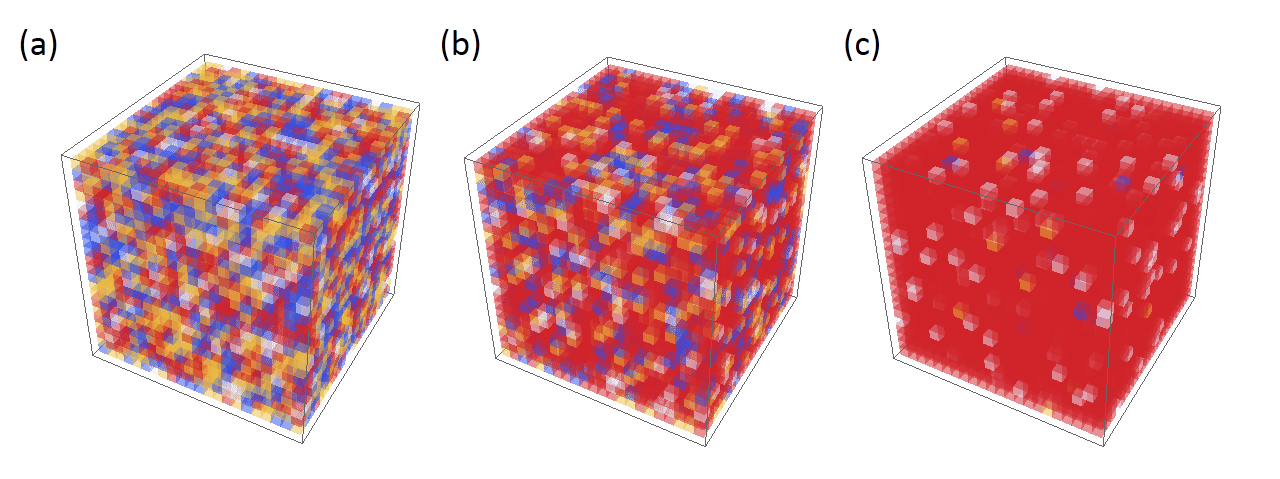}}
  \captionsetup{width=1\textwidth}
  \caption{\small {Percolation of red cluster in a diluted Potts model with constant concentration at a critical point. (a)  At high temperatures three spin states (red, blue, yellow) equally occupy the system. White colors are diluted sites. (b) When temperature decreases, one of the spin states (red here) starts to percolate. (c) At lower temperatures red sites completely occupy the system as a huge cluster.}}\label{Fig.4}
\end{figure}

Hamiltonian of a random site-diluted Potts system is just like Eq. (\ref{eq.2}). Again random dilution is represented by occupation variable $\varepsilon_r=0,1$, but here the probability of occupation for each site is determined by concentration $p\, \,(p=1 \textrm{ means no site is removed})$ \cite{c.20,c.22}. According to C-J conjecture there exists a critical concentration at $0\leq p_t\leq 1$ (Fig. (\ref{Fig.3})) \cite{c.13,c.20,c.22}. For $p\geq p_t$ the phase transition line is $1st$ order which would changes to $2nd$ order at a tri-critical point $p_t$ (Fig. (\ref{Fig.3})). According to this figure, more diluted sites leads to a decrease in critical temperature. When $p$ approaches the critical concentration $p_t$ along the $1st$ order transition line, latent heat and surface tension would vanish \cite{c.13,c.20,c.22}. At this point correlation length diverges as well \cite{c.13,c.20,c.22}. At lower temperatures, $Z_3$ symmetry breaks spontaneously and one of the three spin states starts to percolate and occupies the entire lattice (Fig. (\ref{Fig.4})).

\begin{figure}[t!]
  \centering
  \frame{\includegraphics[width=1\textwidth]{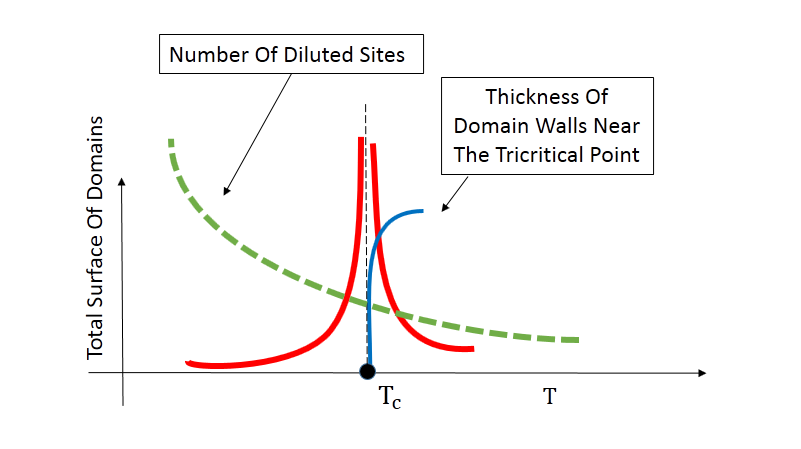}}
  \captionsetup{width=1\textwidth}
  \caption{\small {Behavior of the total clusters walls thickness near a tri-critical point (blue line). Near the tri-critical point total surface area diverges (red line).}}\label{Fig.5}
\end{figure}

We want to study the zero surface tension in more details. As susceptibility diverges at the tri-critical point (see next section), the total surface area of domain clusters diverges too. Near this point the total surface area of domain clusters become infinitely large and according to Eq. (\ref{eq.3}), no matter how large diluted sites number is, the walls thickness approaches zero (Fig. (\ref{Fig.5})). So there exists a critical number of diluted sites $N_{critical}$ for which if $0\leq N \leq N_{critical}$ the equivalent thickness is $0\leq \Delta\ell \leq \Delta\ell_{max}$ and phase transition would be first order. But when $N\rightarrow N_{critical}$ then $\Delta\ell \rightarrow 0$ and a tri-critical phase transition happens (Fig. (\ref{Fig.5})).

We can interpret the zero surface tension in another way. According to surface tension definition

\begin{equation}\label{eq.4}
\sigma = \frac{energy}{area}
\end{equation}

and the fact that energy stored in domain clusters walls is limited, near a tri-critical point the total surface area of all clusters diverges and as a consequence the surface tension approaches zero. It is shown that this surfaces are fractal and unsmooth \cite{c.23,c.24,c.25}. Later we will use this concepts in order to solve the cosmological domain wall problem. In the next section we study the properties of phase transition near a tri-critical point in Landau theory.

\section{Tri-critical Phase Transition in Landau Theory}
\label{sec.3}

In Landau theory, the general form of free energy as a function of order parameter is given by

\begin{equation}\label{eq.5}
\begin{gathered}
G=G_0 + \frac{1}{2} a (T-T_0) Q^2 + \frac{1}{4} b Q^4 + \frac{1}{6} c Q^6 \\
\begin{cases}
b > 0 \quad : \quad Second\, \, Order \\
b < 0 \quad : \quad First\, \, Order \\
b \sim 0 \quad : \quad Tri-critical
\end{cases}
\end{gathered}
\end{equation}

Here the condition $G(Q)=G(-Q)$ requires that the power series only has even powers of $Q$. The minimum of the free energy defines the equilibrium state. For a tri-critical phase transition we have

\begin{equation}\label{eq.6}
G=G_0 + \frac{1}{2} a (T-T_0) Q^2 + \frac{1}{6} c Q^6
\end{equation}

The order parameter is

\begin{equation}\label{eq.7}
\begin{gathered}
G=G_0 + \frac{1}{2} a (T-T_c) Q^2 + \frac{1}{6} c Q^6 \\
\frac{\partial G}{\partial Q} = a (T-T_c) Q + c Q^5 = 0 \\
Q = 0,{\frac{a}{c} (T_c-T)}^\frac{1}{4}
\end{gathered}
\end{equation}

Entropy become

\begin{equation}\label{eq.8}
\begin{gathered}
S = - \frac{\partial G}{\partial T} \\
\Delta S = -\frac{1}{2} a Q^2 = \begin{cases}
\ 0 \qquad & T>T_c \\
\-\frac{1}{2}{\frac{a^3}{c}(T_c-T)}^{\frac{1}{2}} \qquad & T<T_c
 \end{cases}
 \end{gathered}
\end{equation}

\begin{figure}[t!]
  \centering
  \frame{\includegraphics[width=1\textwidth]{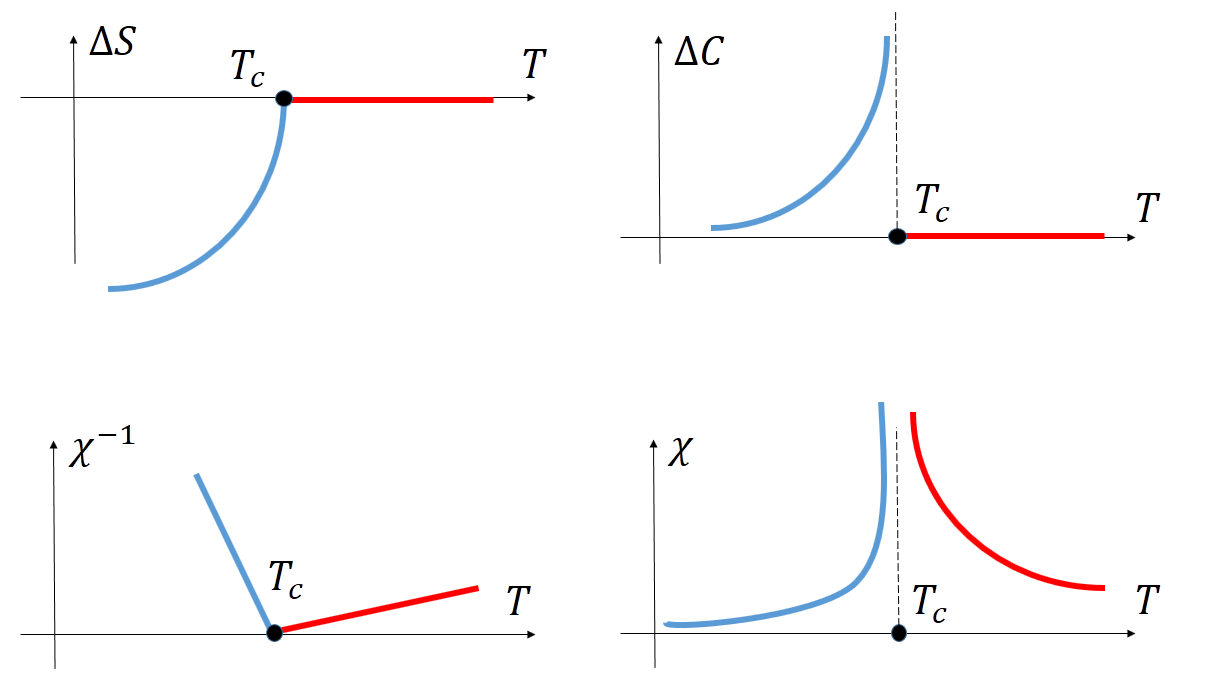}}
  \captionsetup{width=1\textwidth}
  \caption{\small {Behavior of entropy, heat capacity, and susceptibility near the tri-critical point.}}\label{Fig.6}
\end{figure}

For heat capacity and susceptibility we have

\begin{equation}\label{eq.9}
\Delta C = T \frac{\partial \Delta S}{\partial T} = \begin{cases}
\ 0 \qquad & T>T_c \\
\ \frac{1}{4}\sqrt{{\frac{a^3}{c}}} T (T_c-T)^{-\frac{1}{2}} \qquad & T<T_c
 \end{cases}
\end{equation}

\begin{equation}\label{eq.10}
\chi^{-1} = \frac{\partial ^2 G}{\partial T^2} = a |T-T_c| + 5c Q^4= \begin{cases}
\ a(T-T_c) \qquad & T>T_c \\
\ -4a(T-T_c) \qquad & T<T_c
 \end{cases}
\end{equation}

Divergence of susceptibility implies the divergence of correlation length. In the next section, we want to apply these results together with percolation theory and C-J conjecture to present a new mechanism to solve the cosmological domain walls created during the $Z_3$ electroweak symmetry breaking in NMSSM.

\section{NMSSM Electroweak Symmetry Breaking }
\label{sec.4}

Minimal supersymmetric standard model (MSSM) is an extension to the standard model that realizes supersymmetry. Three principle motivations for MSSM are naturalness, gauge coupling unification, and dark matter \cite{c.1,c.2}. It considers only the minimum number of new particle states and new interactions consistent with phenomenology.

Next to minimal supersymmetric standard model (NMSSM) is a supersymmetric extension to standard model that adds an additional singlet chiral superfluid to the MSSM and can be used to dynamically generate the $\mu$ term, solving the $\mu$-problem \cite{c.1,c.2}. The cubic renormalizable superpotential in NMSSM is \cite{c.26}:

\begin{equation}\label{eq.11}
\textit{W}_{ren} = \lambda S H_1 H_2 + \frac{\kappa}{3} S^3 + Y^{(u)} Q U^c H_1 + Y^{(d)} Q D^c H_2 + Y^{(e)} L e^c H_2
\end{equation}

Here $H_1$ and $H_2$ are Higgs fields, $Y$s are Yukawa couplings, and $S$ is massless gauge singlet field coupled to the Higgs field in the first term. The second term explicitly breaks the Peccei - Quinn symmetry, while violating no other wanted symmetry.

According to Kibble-Zurek mechanism \cite{c.3,c.4,c.5}, the $Z_3$ discrete symmetry break during the phase transition associated with NMSSM electroweak symmetry breaking at $\textit{O} (10)^2$ GeV energy scale leads to the formation of domain walls which are topological defects. These walls would form at the boundaries of different degenerate vacua, and due to the existence of causal horizon in an expanding universe these degenerate vacua can't be in causal connection (we will see that there is a possibility for these vacua to be causally connected beyond their horizons). It was first noted by Zel'dovic, Kobzarev and Okun \cite{c.6}, that if discrete symmetries of scalar field theory are spontaneously broken as the universe cools down, there would be further drastic problems in its evolution.

If such domains remain they can over close the universe by dominating the energy density \cite{c.6,c.9,c.27}. It is shown that these defects can cause huge amount of anisotropies in the CMB, and to avoid this problem their energy scale must be less than a few MeV \cite{c.7}. Actually to be more concise it must be less than 0.93 MeV \cite{c.8}. As an indispensable consequence these problematic walls must be lifted before the nucleosynthesis era. First suggested by Zel'dovic \textit{et al.} \cite{c.6}. breaking the degeneracy of the vacua, eventually leads to the dominance of true vacuum.

We want to see how previous discussions can solve the cosmological domain walls problem. The condition for the walls to be unstable and their energy density decays exponentially fast is \cite{c.6,c.9,c.27}:

\begin{equation}\label{eq.12}
\delta E > \frac{\sigma}{R}
\end{equation}

Here $\delta E$ is pressure or difference in energy density between the two minima. $\sigma$ and $R$ are surface energy density and radius curvature of the wall. For the cosmological walls $\sigma \sim \upsilon^3$ where $\upsilon$ is the vacuum expectation value of the field which is of order $10^2$ GeV for the electroweak energy scale.

\subsection{Non-degenerate Vacua}
\label{sec.4.1}

\subsubsection{Uncorrelated Domains}
\label{sec.4.1.1}

Just after the $Z_3$ symmetry breaking, different uncorrelated regions in the early universe randomly choose one of the three vacua with probability 0.3333. According to percolation theory the threshold probability $P_c$ for a random site percolation in three dimensions is 0.3116 \cite{c.10,c.11}.
\begin{figure}[t!]
  \centering
  \frame{\includegraphics[width=1\textwidth]{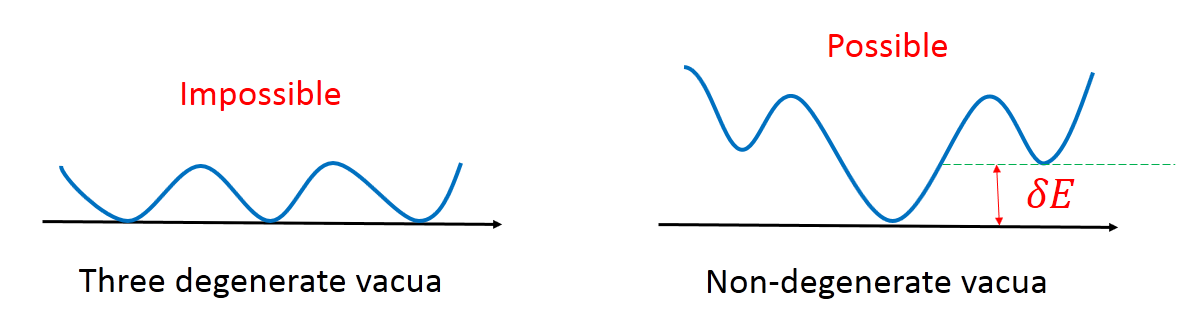}}
  \captionsetup{width=1\textwidth}
  \caption{\small {(Left) Three degenerate vacua. One vacuum is dominated due to the random percolation threshold in three dimensions.}}\label{Fig.7}
\end{figure}
 Due to the fact that $\frac{1}{3}>0.3116$, it is impossible to have three degenerate vacua. In this case, there is always a dominant vacuum which destroys the degeneracy, and if the difference in energy density become larger than surface tension of the wall as stated in Eq. (\ref{eq.12}), then the walls become unstable and decay. As a consequence, there would be no cosmological domain wall problem. In this case (uncorrelated domains), if $\delta E < \frac{\sigma}{R}$  the problem won't be solve.

\subsubsection{Weakly Coupled Domains}
\label{sec.4.1.2}

\begin{figure}[b!]
  \centering
  \frame{\includegraphics[width=1\textwidth]{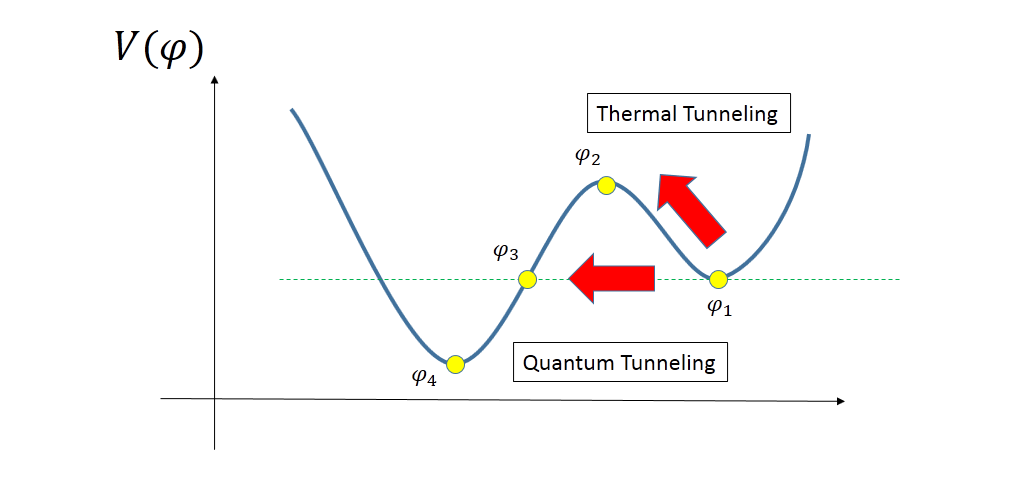}}
  \captionsetup{width=1\textwidth}
  \caption{\small {Quantum and thermal tunneling between two potential minima (figure adopted from \cite{c.28}).}}\label{Fig.8}
\end{figure}

Now we want to consider a situation in which distinct vacua are causally connected beyond their horizons. This connection might be rooted in entanglement or a weakly coupled field which would be correlated on super-horizon scales at the end of inflation (\cite{c.9,c.27} and references in them) or because of quantum and thermal tunneling (Fig. (\ref{Fig.8})). The tunneling rate from false to true vacuum is given by \cite{c.28}:

\begin{equation}\label{eq.13}
\Gamma = A e^{-\frac{B}{\hbar}}
\end{equation}

Here $B$ is the Euclidean action of the field configuration that leads to tunneling and can be calculated via WKB generalization of field theory. We can also calculate $A$ using the path integral approach \cite{c.28}.

\begin{figure}[t!]
  \centering
  \includegraphics[width=1\textwidth]{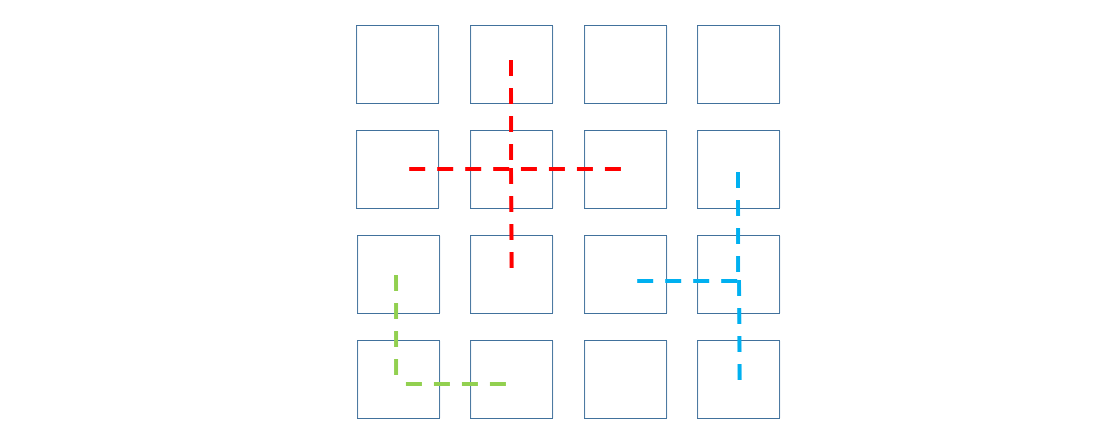}
  \captionsetup{width=1\textwidth}
  \caption{\small {Cosmological domains with causal connections beyond their horizon.}}\label{Fig.9}
\end{figure}

So as we see in Fig. (\ref{Fig.9}), there exists a possibility for the domains of $Z_3$ vacua in NMSSM electroweak symmetry breaking to be causally connected beyond their horizons to their nearest neighbors. Now we can see better why it was claimed that an expanding universe filled with regions of $Z_3$ vacua and domain walls created during the NMSSM electroweak symmetry breaking, belongs to the same universality class of a $3d$ three states diluted Potts model. Cosmological domain walls here play the same role as equivalent domain clusters walls in the diluted Potts model do. As long as each vacuum region could be connected to its nearest neighbors beyond its horizon, the universe expansion won't be problematic for the universality considerations.

\begin{figure}[b!]
  \centering
  \frame{\includegraphics[width=1\textwidth]{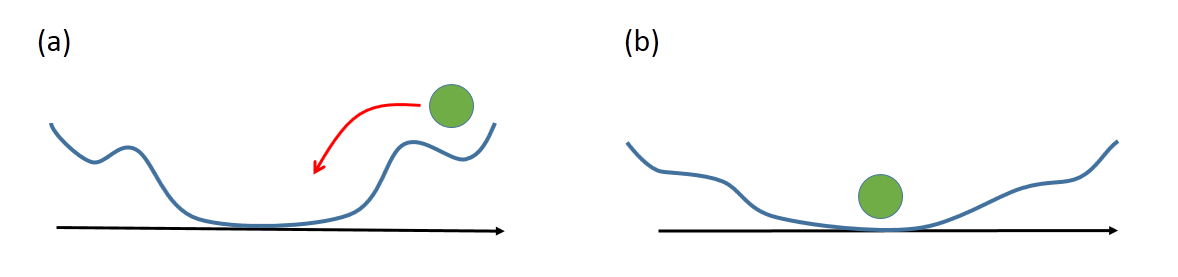}}
  \captionsetup{width=1\textwidth}
  \caption{\small {(Left) The early universe might be trapped in a metastable state for $0 < \delta < \delta_{max}$. (Right) Inflation finishes due to the condition $\delta E >\frac{\sigma}{R}$.}}\label{Fig.10}
\end{figure}

Being in the same universality class, we can use the results and behaviors of the diluted Potts system near its tri-critical point, in order to study the behavior of the early universe filled with $Z_3$ vacua. Moreover, there must be a tri-critical point in the chronology of early universe during the NMSSM electroweak symmetry breaking/phase transition which we are going to study in more details. There exists two different stages in the evolution of the universe:

First, the domain walls thickness is $0 < \delta < \delta_{max}$ which corresponds to $0 < N < N_{critical}$ in the diluted Potts model with equivalent thickness $\Delta\ell$. In this case the universe might be trapped in a metastable state and inflates (Fig. (\ref{Fig.10})). Due to the fact that percolation threshold is less than the distribution probability $(0.3116<1/3)$ there would exists a non-degenerated vacuum. Similar to the completely disconnected case, if $\delta E >\frac{\sigma}{R}$ the walls decay exponentially fast and vanish, which helps the universe to exit out of the inflationary state (Fig. (\ref{Fig.10}.b)).

Second stage would happens if $\delta E <\frac{\sigma}{R}$ in which the domain walls do not decay. In this case, according to Zel'dovic \textit{et al.} \cite{c.6} the early universe would experiences an expansion due to energy density of the walls which might become dangerous if their growth lasts for a long time. But based on the results of diluted Potts model, and the fact of being in the same universality class, near the cosmological tri-critical point $\delta \rightarrow 0$ that corresponds to $N\rightarrow N_{critical}(\delta \ell \rightarrow 0)$ in the diluted Potts model. This means that the cosmological walls expansion should be ceased as the universe approaches tri-critical point. We can interpret this by the fact that near the tri-critical point, total cluster surfaces is fractal and diverges, so we have

\begin{equation}\label{eq.14}
Thickness \propto \frac{Volume}{Area\rightarrow \infty} \sim 0
\end{equation}

There is another reason for zero domain wall thickness presented in subsection (\ref{sec.4.3}). In the next three subsections, we want to study the physical consequences of correlation length divergence, zero surface tension, and fluctuations of all length scales near the tri-critical point.

\begin{figure}[t!]
  \centering
  \includegraphics[width=1\textwidth]{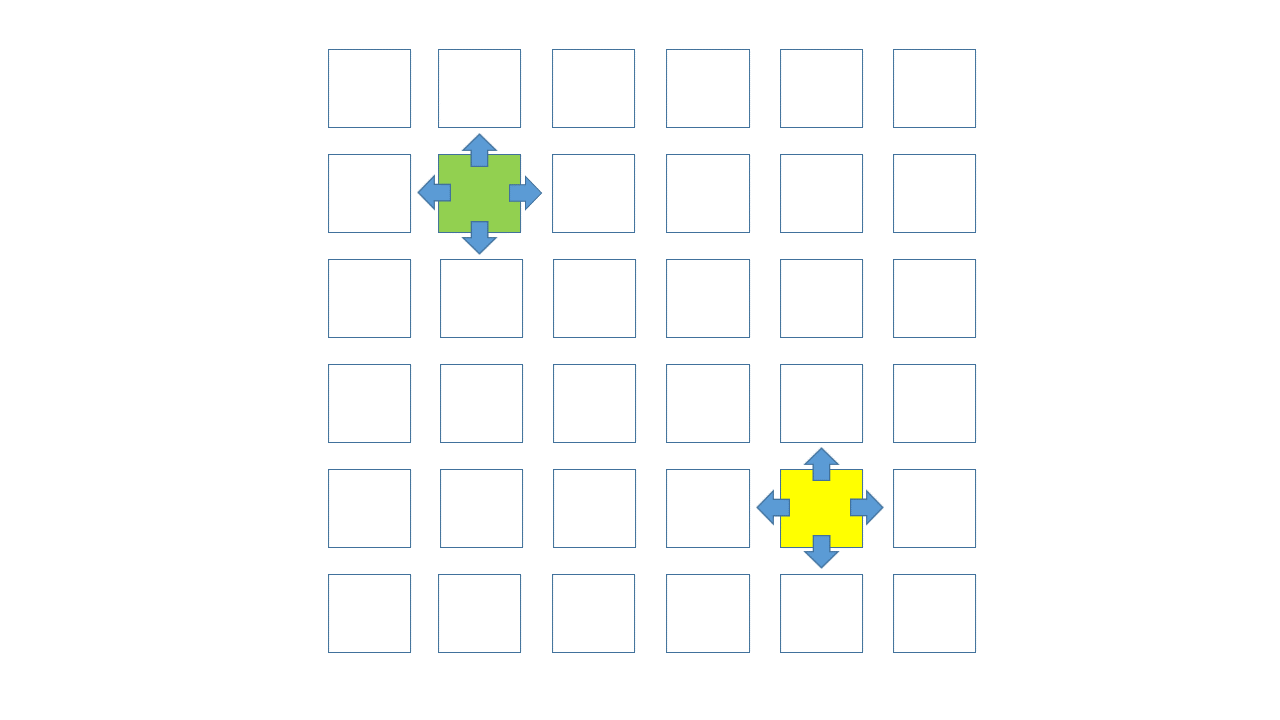}
  \captionsetup{width=1\textwidth}
  \caption{\small {Two completely disconnected $Z_3$ vacua domains at different parts of the early universe (horizon problem) can be connected at the tri-critical point.}}\label{Fig.11}
\end{figure}

\subsection{Electroweak Horizon Problem}
\label{sec.4.2}

A universe randomly filled with three different $Z_3$ vacua is completely inhomogeneous. In this subsection, we want to use the idea of causally connected neighbors to see how NMSSM electroweak symmetry breaking horizon problem could be solved. Fig. (\ref{Fig.11}) shows two completely disconnected and different $Z_3$ vacua at different parts of the universe. No matter how far they are, these two domains could have causal connection with each other at the tri-critical point.  The only constraint here is that these $Z_3$ vacua can be causally connected to their nearest neighbors beyond their horizon.

\begin{figure}[t!]
  \centering
  \includegraphics[width=1\textwidth]{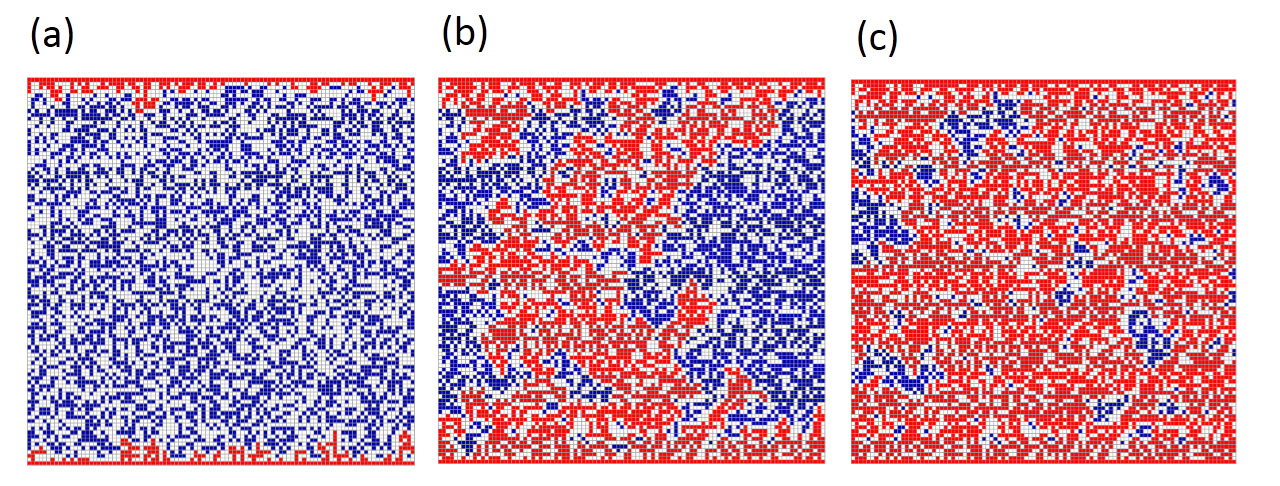}
  \captionsetup{width=1\textwidth}
  \caption{\small {Emergence of a huge spanning cluster at a critical point in a spin system. Each site in this figure represents one of the two (up and down) states in the spin system. (a) Two equally distributed spin states at high temperatures. (b) At lower temperatures, near the critical point a huge blue cluster emerges that connects two sides of the system (for clarity this cluster is represented by red color). (c) At lower temperatures this cluster span through the entire lattice (figure from "Percolation on a Square Grid" from the Wolfram Demonstrations Project).}}\label{Fig.12}
\end{figure}

In order to see how divergence of correlation length can solve the electroweak horizon problem, we consider behavior of a two spin states Potts model near its critical point. We choose this example for simplicity and clarity, the results is the same for the $3d$ three states diluted Potts model and the early universe filled with $Z_3$ vacua near their tri-critical point where correlation length diverges. As shown in Fig. (\ref{Fig.12}), blue and white sites represent two spin states (up and down) in the spin system. At high temperatures, there are equal number of blue and white sites (zero magnetization) that according to the Potts Hamiltonian can only have causal connection with their nearest neighbors (Fig. (\ref{Fig.12}.a)). When temperature decreases, near the critical point up-down symmetry breaks spontaneously and suddenly a huge spanning domain cluster emerges that connects the different edges of the system (Fig. (\ref{Fig.12}.b)) (for clarity this huge cluster is represented by red color). At this point due to the divergence of correlation length, the two previously disconnected sites at different parts of the system/universe in Fig. (\ref{Fig.11}) are now in the same huge cluster with identical properties and a causal connection. When temperature decreases more this spin cluster completely occupies the entire system (Fig. (\ref{Fig.12}.c)) and the result would be a homogeneous system.

This is just like what happens in the $3d$ three states diluted Potts model (Fig.(\ref{Fig.4})) and the universe filled with $Z_3$ vacua which are connected via a weakly coupled field. For these systems, near the tri-critical point correlation length diverges and there would exists a huge spanning and dominant domain cluster that completely occupies the entire system/universe after the phase transition.

With this mechanism we can avoid the NMSSM electroweak horizon problem and explain the homogeneity of universe (one dominant $Z_3$ vacuum rather that three) at the end of electroweak epoch. In this scenario, causally disconnected patches of the early universe at different parts that are completely far from each other can be aware of themselves at the tri-critical point. So as we see, the simple hypothesis of a weakly coupled field that locally connects the nearest neighbors leads to a global causal connection at the tri-critical point.

\subsection{Surface Energy Density}
\label{sec.4.3}

The surface tension or surface energy density of a $3d$ three states diluted Potts model near a tri-critical point approaches zero \cite{c.13,c.20,c.22}. In section (\ref{sec.2}), we interpreted this as a consequence of fractal cluster surfaces which leads to the divergence of surface area of all clusters and zero equivalent thickness $\delta \ell$. According to be in the same universality class, the surface tension of the cosmological domain walls must approaches zero at its tri-critical point too.

\begin{figure}[t!]
  \centering
  \frame{\includegraphics[width=1\textwidth]{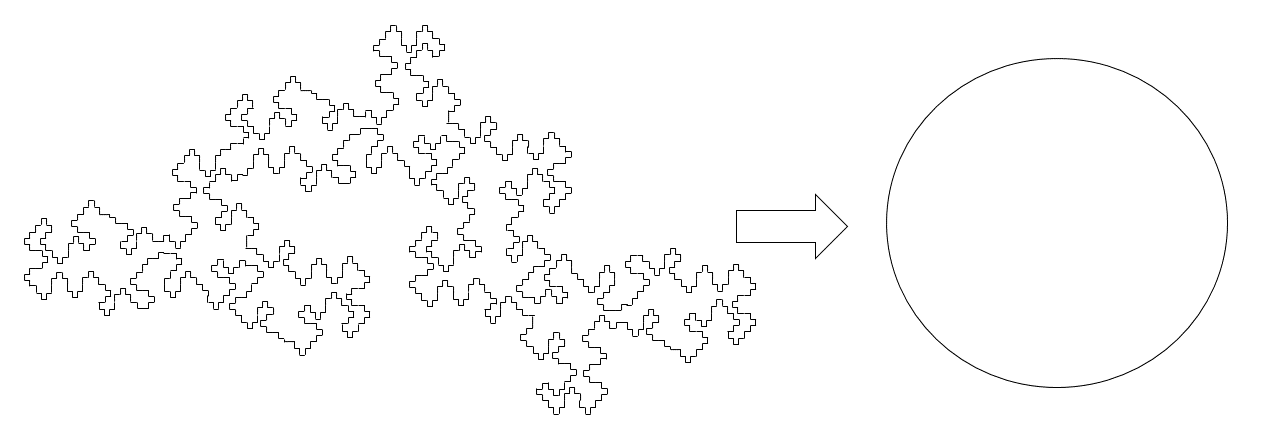}}
  \captionsetup{width=1\textwidth}
  \caption{\small {Near the tri-critical point, surface of domain walls is fractal which is highly unsmooth and unfavorable. The surface uses its energy to minimize its surface area.}}\label{Fig.13}
\end{figure}

For a cosmological domain wall, zero surface tension means no repulsive gravitational field \cite{c.7,c.29,c.30}. Near this point according to Eq. (\ref{eq.12}) the pressure term dominates the tension and cosmological domain walls should decay exponentially fast.  Moreover, for the surface tension $\sigma$ we have

\begin{equation}\label{eq.15}
\sigma = \varrho_{\nu} \, \delta
\end{equation}

Here $\varrho_{\nu}$ is vacuum energy density and $\delta$ is the wall thickness. As we see, zero surface tension leads to zero thickness of domain walls, which again can be interpreted as a direct consequence of the fractal nature of cosmological cluster surfaces near the tri-critical point.

Fractal surfaces of these high energy domain walls are extraordinary unsmooth and unfavorable. So locally at each point, the wall uses its internal energy to minimize its surface area. As it is shown in Fig. (\ref{Fig.14}) , near a tri-critical point some interesting things happen:

\begin{figure}[b!]
  \centering
  \frame{\includegraphics[width=1\textwidth]{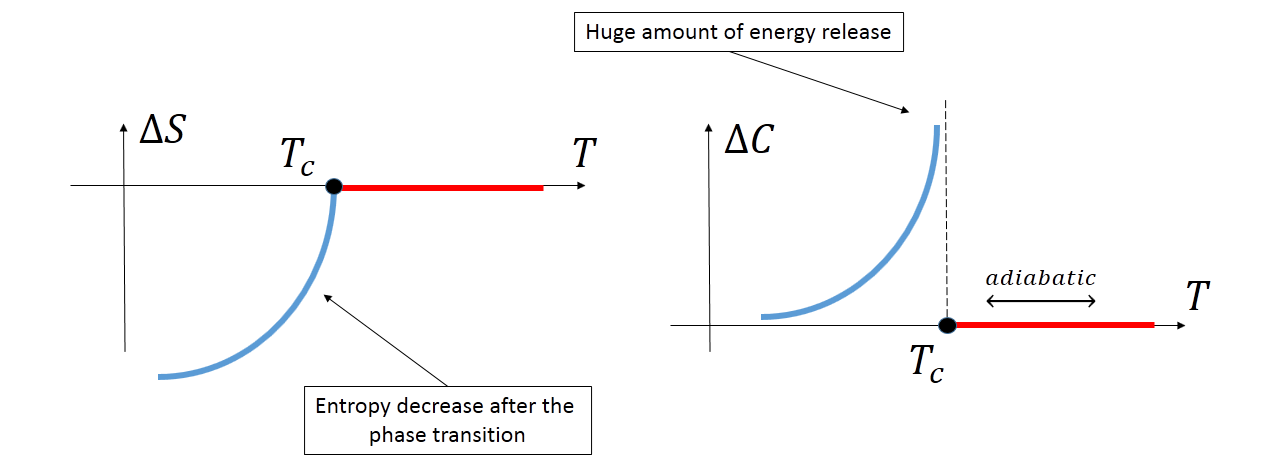}}
  \captionsetup{width=1\textwidth}
  \caption{\small {Entropy decrease after the tri-critical point can offers a solution to the low entropy problem of the early universe.}}\label{Fig.14}
\end{figure}

No matter prior to the tri-critical point we have inflation or not, there is a huge amount of energy release at constant temperature at this point due to the divergence of heat capacity which might be used in the baryogenesis mechanism. For temperatures higher than tri-critical point the heat capacity is zero which indicates an adiabatic process, so prior to this point we have adiabatic fluctuations. Near the tri-critical point due to the divergence of correlation length there would exist adiabatic fluctuations with different length scales from zero to the size of universe.

One of the most interesting features of the tri-critical point is an entropy decrease after the phase transition. This might explains the initial low entropy condition of the early universe \cite{c.31,c.32}. Domain walls can also sweep the GUT monopoles and vanish them completely; the result would be unstable walls \cite{c.33}.

\subsection{Density Perturbation}
\label{sec.4.4}

In order to dilute the abundance of thermal and non-thermal relics, the idea of a period of late-time inflation at electroweak energy scale or lower, has been suggested in \cite{c.14,c.15,c.16,c.17,c.18}. This late-time inflation would be problematic for primordial density fluctuations produced by ordinary GUT inflation. There are two different approaches toward this problem:

First \cite{c.14,c.15}, it is believed that a weak scale inflation can't give rises to cosmological density perturbation of the magnitude required for large scale structure formation. In this scenario, for the late-time inflation to not erase the primordial fluctuations, the number of e-folding must be sufficiently small $(\sim 10)$, therefore primordial density perturbations could be preserved on large scales. Second \cite{c.16,c.17}, there are some supersymmetric inflationary model, based on new space dimensions or allowing more than one field relevant for inflation, in which acceptable density perturbation required for large scale structure can be generated. Whether an NMSSM electroweak symmetry breaking mechanism disturb the fluctuations of GUT inflation or not, the scenario presented in this paper can solve the problem by considering the fact that near the tri-critical point there are adiabatic fluctuations of all length scales in the universe which can be responsible for the further cosmological structure formations.

\subsection{Some Comments}
\label{sec.4.5}

The mechanism presented in this paper is in accordance with the standard cosmology, i.e. the NMSSM electroweak epoch comes after the ordinary inflation (Fig. (\ref{Fig.15}). Along with this standard framework, there are some cosmologist who place the electroweak epoch together with ordinary inflation after the GUT epoch \cite{c.34,c.35}. Also, recently there have been some attempts to study inflation, as a direct consequence of the NMSSM mechanism \cite{c.36}. Based on our results in this paper, we suggest that the ordinary inflation itself might be a consequence of NMSSM electroweak symmetry breaking/phase transition, from the tri-critical point of view.

There exists a difficulty to accomplish this task: in order to make a connection between the Potts model and the early universe filled with $Z_3$ vacua created during the NMSSM electroweak symmetry breaking, we assumed that there exists a causal connection between the nearest vacua beyond their horizons via a weakly coupled field which is rooted in a prior inflationary epoch \cite{c.9,c.27}. If we want the ordinary inflation to be a consequence of the NMSSM electroweak symmetry breaking/phase transition, then we must found a mechanism to causally connects the different $Z_3$ vacua beyond their horizons. If this condition is met then:

\begin{figure}[t!]
  \centering
  \includegraphics[width=1\textwidth]{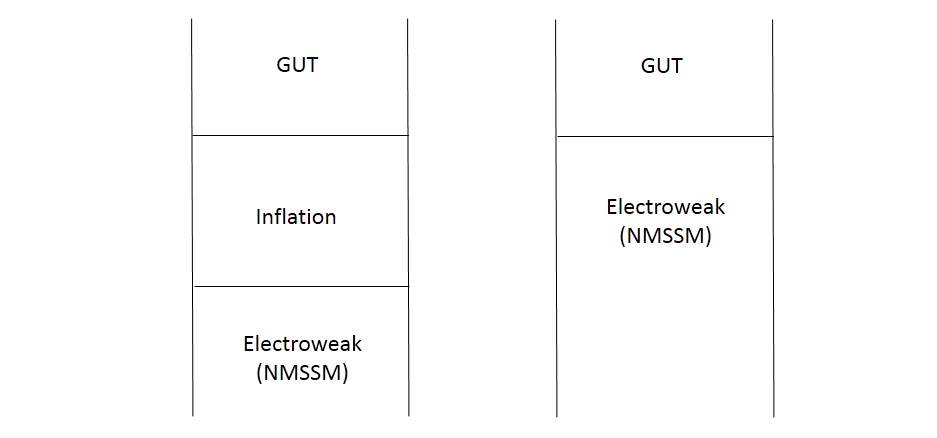}
  \captionsetup{width=1\textwidth}
  \caption{\small {(Left) The standard chronology of the early universe to which this paper is based   on. (Right) The less standard cosmology.}}\label{Fig.15}
\end{figure}

In order to solve the flatness problem, a trapped universe in the metastable state prior to the tri-critical point, supercools and experiences an inflationary state. This idea resembles the old inflation scenario \cite{c.37}, but here the difference is that it has a graceful exit due to approaching the tri-critical point, or walls decay before reaching this point. The huge amount of energy release at the tri-critical point can be responsible for the ordinary reheating process.

In order to solve the horizon problem, as we see in previous subsections, we have percolation mechanism and existence of a spanning huge cluster at the tri-critical point which connects different parts of the universe. The advantage of this mechanism is that unlike the other solutions to the horizon problem, this mechanism doesn't lead to a multiverse scenario.

In order to obtain the proper density fluctuation needed for further structure formation, along with the fact that primordial quantum fluctuations can be enlarged up to cosmic scales while the universe is trapped in a pre tri-critical metastable state, there exist adiabatic fluctuations at the tri-critical point with different length scales from zero to infinity which leads to further structure formations.

In order to solve the GUT monopoles problem, along with an inflation due to the metastable state, domain walls can sweep them too \cite{c.33}.

\section{Pure Gluodynamics and Center Domains}
\label{sec.5}

In this section, we want to study the thickness of domain walls created during the $Z_3$ center symmetry breaking in a pure SU(3) theory and its application in early cosmology. We will see that there might exists a tri-critical point at $\textit{O} (10)^2$ MeV in the evolution of early universe which is different from the NMSSM tri-critical point at $\textit{O} (10)^2$ GeV.

When a discrete symmetry is spontaneously broken, different phases exist that are distinguished by some order parameter. Different regions of space are in different phases. As we said in the previous sections these regions are separated by domain walls with their free energy proportional to surface area. When we neglect quarks as in quenched (static quarks) approximation of QCD, the $Z_3$ center symmetry of SU(3) gauge group play a crucial rule. The order parameter in pure SU(3) is Polyakov loop \cite{c.38}

\begin{equation}\label{eq.16}
 L=\frac{1}{3}\, Tr\, P \exp{\left [ig\, \int_{0}^{1/T} A_4(\tau ,x)\, d\tau \right ]}
\end{equation}

Here $P$ is ordering operator, $\tau $ is imaginary time and $ A_4=iA^{0a}=iA^{0a}\, \lambda ^{0a}/2 $. The thermal average of Polyakov loop distinguishes the quark gluon plasma from hadron phases. $\langle L(x)\rangle$ vanishes in the confined phase. In the deconfined phase it takes the value of the three elements of $Z_3$, the center of SU(3): $\, \, \exp(\frac{2i\pi \nu }{3})$ with $\nu=0,1,2$ \cite{c.38}. These three values construct center domains which play a crucial rule in describing the properties of strongly interacting quark gluon plasma (sQGP) produced in RHIC and LHC such as near perfect fluidity and strong jet quenching \cite{c.39}.

If we apply C-J conjecture to center domains, the result would be a pseudo-inflation. By pseudo we mean without an inflationary era (exponential increase of scale factor $a(t)$), there would be an energy release together with subsequent injected entropy that can dilute the abundance of thermal and non-thermal relics of earlier times \cite{c.14,c.15,c.16,c.17,c.18}. For further procedure we need to combine C-J with Y-S conjecture.

\begin{figure}[t!]
  \centering
  \includegraphics[width=1\textwidth]{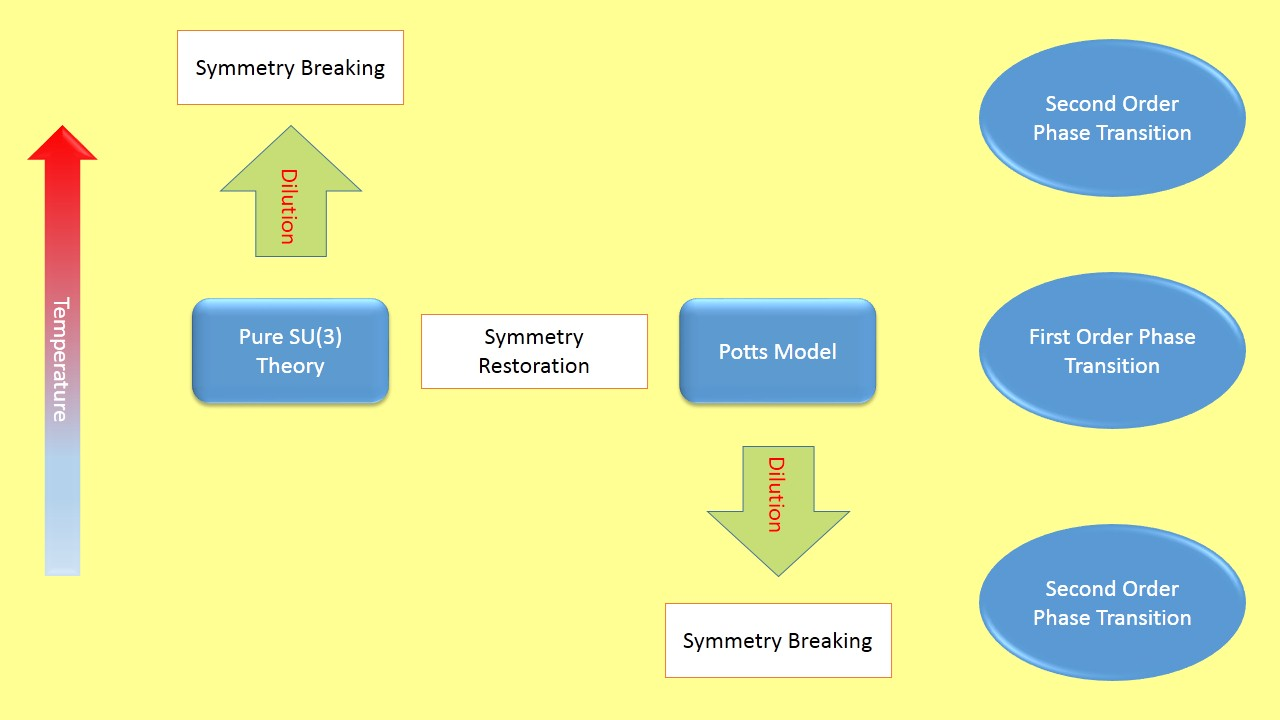}
  \captionsetup{width=1\textwidth}
  \caption{\small {Symmetry between Potts model and SU(3) theory according to Y-S conjecture.}}\label{Fig.16}
\end{figure}

\subsection{Yaffe-Svetitsky Conjecture}
\label{sec.5.1}

According to Y-S conjecture a $(d+1)$ dimensional pure gauge theory undergoing a continuous deconfinement transition is in the same universality class as a $d$-dimensional statistical model with order parameter taking values in the center of the gauge group \cite{c.19,c.40}, i.e. in order to study the $Z_3$ center symmetry breaking and phase transition of a pure SU(3) theory in four dimensions we have to look at a three dimensional three states Potts model.

Earlier in this paper we investigated the properties of a diluted spin systems under C-J conjecture. The question here is that according to Y-S conjecture, do we have the same procedure for a pure SU(3) theory in $4d$? If yes, what would be the SU(3) equivalent of the Potts model diluted sites?

The answer is yes. The idea of diluting sites in a pure SU(3) theory in lattice simulations has been suggested in \cite{c.23,c.41,c.42}. For each spatial point $x$, the authors assigned a sector number $n(x)={1,0,-1}$ in the following manner

\begin{equation}\label{eq.17}
n(x) =
  \begin{cases}
   \ +1 & \text{for } \theta (x) \in [\frac{\pi}{3} + \delta , \pi - \delta ] \\
   \quad 0       & \text{for } \theta (x) \in [-\frac{\pi}{3} + \delta , \frac{\pi}{3} - \delta ]\\
   \, \, -1 & \text{for } \theta (x) \in [-\pi + \delta , -\frac{\pi}{3} - \delta ]
     \end{cases}
      \quad \delta = \frac{\pi}{3}.f
\end{equation}

\begin{figure}[t!]
  \centering
  \includegraphics[width=1\textwidth]{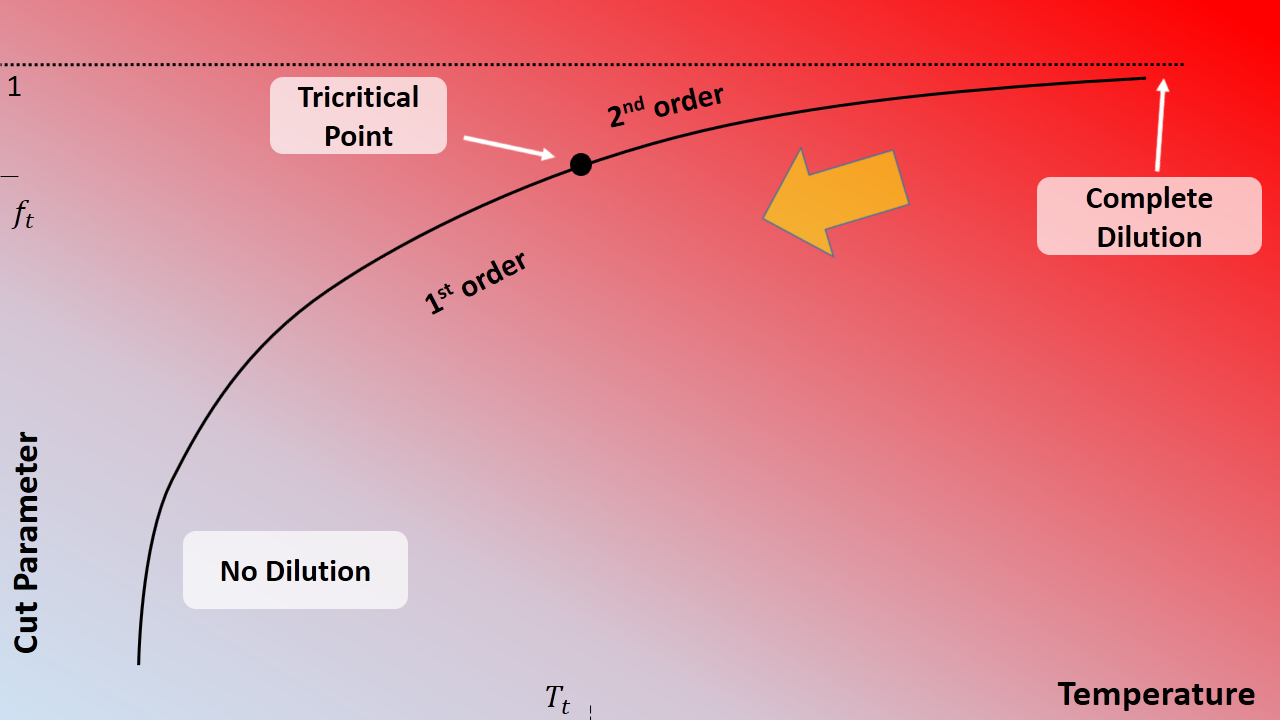}
  \captionsetup{width=1\textwidth}
  \caption{\small {Schematic phase transition line for a diluted pure SU(3) theory in $4d$. Here, in contrast to the Potts model, when temperature decreases the transition line changes from $2nd$ to $1st$ order at a tri-critical point.}}\label{Fig.17}
\end{figure}

Here $\theta(x)$ is the value of local phase of Polyakov loop at site $x$ and $f\in[0,1)$ is cut parameter, which removes undecided sites, i.e. those that their phases $\theta(x)$ is far from each three center angels $0,\pm\frac{2\pi}{3}$ and no sector number is assigned to them \cite{c.23,c.41,c.42}. Here $f=0$ means no site is removed and if $f\rightarrow 1$ all sites are excluded. Two neighboring points $x$ and $y$ are assigned to the same cluster if $n(x)=n(y)$ \cite{c.23,c.41,c.42}. As we see here, the cut parameter in a $4d$ pure SU(3) theory plays the same role as concentration $p$ in the Potts model does. We can now apply C-J conjecture to this theory, i.e. there must exists a critical cut parameter $f_t$ where the phase transition changes from second to first order. As we see in Fig. (\ref{Fig.16})  the only difference between these two theory is that for a pure SU(3) theory, in contrast to the Potts model, the spontaneous symmetry breaking occurs at high temperatures.

According to Y-S conjecture the behavior of these two systems must be the same, or in other words they must belong to the same universality class. The thing we have done here is to extend Y-S conjecture to a diluted version. As we see in Fig. (\ref{Fig.16}) and Fig. (\ref{Fig.17}), everything must be inversed for a diluted pure SU(3) theory in early universe at $\textit{O} (10)^2$ MeV. At high temperatures the transition line for a $4d$ diluted pure SU(3) is $2nd$ order. When temperature decreases, at a critical cut parameter $f_t$ the transition line changes to $1st$ order at a tri-critical point. Unlike the Potts model, for diluted pure SU(3) phase transition near the tri-critical point, the order parameter is non-zero at high temperatures and zero when temperature decreases.

\begin{figure}[t!]
  \centering
  \frame{\includegraphics[width=1\textwidth]{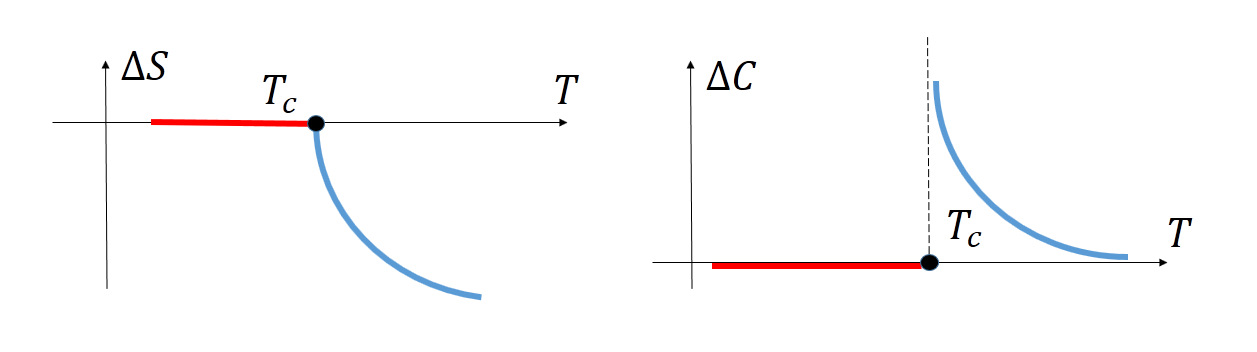}}
  \captionsetup{width=1\textwidth}
  \caption{\small {Inversed Landau theory of phase transition for pure gluodynamics near a tri-critical point.}}\label{Fig.18}
\end{figure}

For the Landau theory to be consistent with diluted pure SU(3) symmetry breaking and phase transition, the behavior of theory must be inversed (Fig. (\ref{Fig.18}), compare it with Fig. (\ref{Fig.6}) and Fig. (\ref{Fig.12})). As temperature decreases and approaches $T_c$, the heat capacity gradually increases which means more energy is released at constant temperature. Moreover, at the same time entropy increases too. Here the increase in heat capacity and entropy occurs without any inflationary and reheating mechanism. This phenomenon is just the consequence of C-J and Y-S conjectures along with the tri-critical phase transition in Landau theory. We must pay attention that this mechanism is for a pure gluodynamics theory ,and due to the fact of being in the same universality class of the Potts model, there is no need to know about details of interactions at the tri-critical point.

The energy release and entropy injection can dilutes the abundance of thermal and non-thermal relics produced at earlier stages. This provides an eligible mechanism for late-time inflation scenario in Ref. \cite{c.14,c.15,c.16,c.17,c.18} without any exponential increase in scale factor $a(t)$ and further reheating process. In this scenario, we can also avoid the problem of large initial baryon chemical potential in Ref. \cite{c.44}. Therein the authors, in order to induce a late-time inflation proposed a scenario in which the universe starts out at a large initial baryon chemical potential condition, which might be against standard cosmological observation.

\section{Conclusion}
\label{sec.6}

In this paper, we presented a new mechanism to connect condensed matter, cosmology, and particle physics. To solve the cosmological domain walls problem created during the NMSSM electroweak symmetry breaking, we studied the conditions in which we could bring a universe filled with electroweak $Z_3$ vacua and a $3d$ three states diluted Potts model together in the same universality class in order to use the results of C-J conjecture, such as zero surface tension and divergence of correlation length and heat capacity. We also combined C-J and Y-S conjectures in order to solve the problem of thermal and non-thermal overproduction problem via a late time pseudo-inflation.

\end{document}